\documentstyle[12pt]{article}

\def\be{\begin{equation}}
\def\ee{\end{equation}}
\def\ba{\begin{array}{c}}
\def\ea{\end{array}}

\def\ben{$$}
 \def\een{$$}

\begin{document}

\titlepage
\vspace*{4cm}

\begin{center}{\Large \bf
${\cal PT}-$symetric regularization

and the new shape-invariant potentials
 }\end{center}

\vspace{5mm}

\begin{center}
Miloslav Znojil
\vspace{3mm}

\'{U}stav jadern\'e fyziky AV \v{C}R, 250 68 \v{R}e\v{z},
Czech Republic\\

homepage: http://gemma.ujf.cas.cz/\~\,znojil

e-mail: znojil@ujf.cas.cz

\end{center}

\vspace{5mm}

\section*{Abstract}

All the well known exactly solvable $s-$wave potentials $V(r),\
r \in (0, \infty)$ may be ${\cal PT}-$symmetrically regularized
(say, by a ``small" imaginary shift of the coordinate axis) and
re-defined as acting on the whole real line, $r \to x-
i\,\varepsilon$, $x \in (-\infty,\infty)$.  There is no surprise
that the new models are still exactly solvable and that their
energies remain real.  What {\em is} an unexpected result?  The
observed drastic and unpredictable changes in {\em the forms} of
the new spectra themselves.

\vspace{9mm}

\noindent
 PACS 03.65.Ge,
03.65.Fd

\vspace{9mm}

\begin{center}
 {\small text of seminar proposed for ATOMKI, Debrecen;

 \today, Latex file debrec.tex }
\end{center}

\newpage

\section{Introduction}

This lecture is {\em mainly} intended as a preliminary review and
summary of the new exactly solvable models obtained by the author,
step-by-step, during the recent months, and accessible already in
the form of several short notes.

In essence, the overall structure or ``contents" of this text on
{\em non-Hermitian} models might just parallel the page 297 of the
review paper \cite{Khare}. There, the exactly solvable {\em
Hermitian} models have been listed and characterized by their so
called shape invariance.

The quoted list of potentials further decays in the two separate
families. Let us call them ${\cal L}$ and ${\cal J}$ for our
present purposes. They are characterized by the respective
Laguerre and Jacobi polynomial forms of their bound states.

In the former family the first subcategory ${\cal L} X$ is
defined on the whole axis, $ x \in (-\infty,\infty)$. It
contains just the quadratic oscillator $V^{(QO)}(x)$ and the
exponential Morse interaction $V^{(EM)}(x)$.

The second subcategory ${\cal L} R$ of the former family
consists of the two best known three-dimensional solvable
models, viz., of the Coulomb or Kepler $V^{(CK)}(r)$ and the
spherical harmonic $V^{(SH)}(r)$, both with $r \in (0, \infty)$
of course.

The latter family ${\cal J}$ has been conveniently split in the
three subcategories with two elements each. These subcategories
are characterized by their ranges of coordinates.

A particle ``lives" on a finite interval in the first
subcategory ${\cal J} Y$.  Temporarily, let us skip and ignore
this case completely.

The second subcategory ${\cal J} X$ is an opposite extreme with
coordinates covering the whole real axis,  $x \in I\!\!R$.  We
may refer to its elements as Rosen Morse $V^{(RM)}(x)$ and the
scarf potential $V^{(SP)}(x)$.

This agrees with the convention accepted in ref. \cite{Khare}.
Similarly, the respective P\"{o}schl-Teller and Eckart forces
$V^{(PT)}(x)$ and $V^{(EF)}(x)$ are simply determined as contained
in the last, third subcategory ${\cal J} R$ of the whole list.

In what follows, I shall try to parallel this classification of
the Hermitian solvable models within the slightly more general,
non-Hermitian framework of the so called ${\cal PT}$ symmetric
quantum mechanics \cite{Bender}.  In detail I shall only outline
some of the most interesting observations, new results and open
questions concerning these brand-new and still fairly unusual
models.

A unifying principle of all the present generalizations will be
the condition of the so called ${\cal PT}$ symmetry \cite{Bender}.
This, in essence, means the commutativity of our Hamiltonians $H$
with the product of the parity ${\cal P}$ and the operator ${\cal
T}$ of time reversal.  One has to note that in all the present
time-independent models the latter operator symbolizes just a
reflection ${\cal T}\, i = -i\, {\cal T}$ of the complex plane
with respect to its real axis.

\section{Various types of the ${\cal PT}-$symmetric potentials}

One of the first studies of the ${\cal PT}$ symmetric potentials
\cite{BG} paid attention to the quartic (i.e., unsolvable)
power-law forces and has been motivated by their close formal
(and, first of all, Fourier-transformation) connection to their
Hermitian power-law analogues.

Similar questions can be, of course, much more easily studied
within the domain of solvable interactions.  Unfortunately,
their explicit descriptions and/or classifications has only been
initiated, quite recently, after a few parallel pioneering
proposals of the ${\cal PT}$ symmetric quantum mechanics by
Andrianov et al \cite{Andrianov}, Bender et al \cite{Bender},
Cannata et al \cite{Cannata} et al.

\subsection{Generalizations within the family ${\cal L}$}

\subsubsection{${\cal L} X$({\cal PT})}

I shall omit here the detailed analysis of the ${\cal PT}$
symmetrized QO and EM models.  My main reason is an idea of
keeping this text as short as reasonable, supported by another
important fact that the majority of the most relevant comments on
the ${\cal PT}$ symmetrized category ${\cal L} X ({\cal PT})$ (and
on its two models denoted, naturally, as $V^{(PTQO)}(x)$ and
$V^{(PTEM)}(x)$) are already easily accessible in the respective
publications \cite{BB} and \cite{Morse}.

\subsubsection{${\cal L} R({\cal PT})$}

I have mixed reasons why I shall also skip the other two
Laguerre-related models $V^{(PTCK)}(x)$ and $V^{(PTSH)}(x)$.

The reason remains similar in the latter case only. Indeed, most
of the facts I know about it did already appear in my letter
\cite{PTHO}.  In contrast to that, I (and, apparently, most of us)
do not know virtually anything about the former model
$V^{(PTCK)}(x)$ yet.  Only some indirect hints have appeared in a
few loosely related publications \cite{coho}. I strongly believe
that the situation will significantly improve soon \cite{LevaiZ}.

\subsection{Generalizations within the subcategory ${\cal J}$ X}

In the frame of our present denotation, the respective ${\cal PT}$
symmetrized models $V^{(PTSP)}(x)$ and $V^{(PTRM)}(x)$ have been
proposed in the recent short communications \cite{Bagchi} and
\cite{shapin}.  Considerations based on the economy of space force
me again to restrain from any re-copied formulae and repeated
comments.

Let me confirm this rule of conduct by one slightly discomforting
exception. One can really feel puzzled by the latter
innocent-looking, smooth and  asymptotically vanishing potential
which can be always uniformly bounded, $|V^{(PTRM)}(x)| <\delta $.
After its deeper and constructive study we were still not that
much surprised by its ability of supporting a ground state in the
weak coupling regime where, by our assumption, both its couplings
were small of the order ${\cal O}(\delta)$ at least.

Nevertheless, one cannot help feeling caught our of guard by the
observation that the related {ground state energy} may become also
{\em arbitrarily large}, $E_0 = {\cal O}(1/\delta^2)$
\cite{shapin}. Similar paradoxes offer one of the reasons why the
solvable models of the ${\cal PT}$ symmetric quantum mechanics
\cite{Bender} deserve a thorough and exhaustive analysis.

\subsection{Generalizations within the subcategory ${\cal J}$ R}

Moving to a central point of this paper, let us point out that
there exists one key difference between the previous oscillators
and the ``last two" ${\cal PT}$ symmetrized models
$V^{(PTEF)}(r)$ and $V^{(PTPT)}(r)$ since both the Hermitian
predecessors of these two forces are, generically, strongly
singular in the origin.

In the former model $V^{(PTEF)}(r)$, its smoothness or
regularization has been achieved by a small, local deformation
of the integration path $r = r(t)$, $t \in I\!\!R$ near the
origin.  In addition, a complex rotation of one of the couplings
by $\pi/2$ has been employed, in our preprint \cite{Ecka}, as a
guarantee of the ${\cal PT}$ symmetry.

Such a complexification changes the spectrum and becomes a
source of its several unusual features.  A detailed inspection
reveals, e.g., that an increase of the repulsion can {\em lower}
(!) the energy. More details will be recollected here in sect.
\ref{sEcka} below.

In a less drastic approach to the singular forces we preserved
the real values of the couplings in $V^{(PTPT)}(r)$. We employed
the ``minimal" ${\cal PT}$ symmetrization $r(t) = t -
i\,\varepsilon$ of refs. \cite{BG} or \cite{PTHO}. Detailed
properties of the resulting ``last" solvable model
have been described in the preprint \cite{PTP}.
At length they will also be discussed in sect. \ref{ssPTP} below.

Quotation marks in the word ``last" mean that we introduced one
more model very recently \cite{hulth}.  The resulting force
$V^{(PTHS)}(x)$ of a Hulth\'en-type shape fits nicely in the
following scheme,
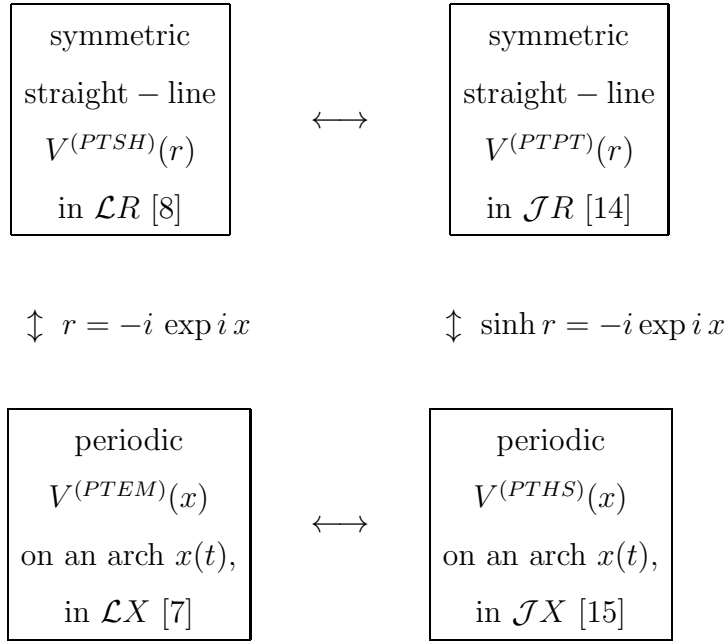
\ben
\ba \\
\begin{array}{|c|} \hline
{\rm symmetric}\\
{\rm straight-line}\\
V^{(PTSH)}(r)
\\
{\rm in\ }{\cal L} R \ \cite{PTHO}\\
\hline \ea \
\ \ \ \ \ \ \stackrel{}{ \longleftrightarrow } \ \ \ \
\ \ \ \begin{array}{|c|} \hline
{\rm symmetric}\\
{\rm straight-line}\\
V^{(PTPT)}(r)
\\
{\rm in\ }{\cal J} R \
\cite{PTP}\\
\hline \ea
\\
\\ \ \ \ \  \  \ \ \ \
  \updownarrow \ r = -i\, \exp i\,x  {\rm }\
 \ \ \  \ \ \ \  \ \ \ \ \
\ \ \ \ \  \updownarrow \
 \sinh r= - i \exp {i\,x}
   {\rm }\  \ \\
\\
\begin{array}{|c|} \hline
{\rm periodic}\\
V^{(PTEM)}(x)
\\
{\rm on \ an \ arch\ }x(t),\\ {\rm in\ }{\cal L} X \
\cite{Morse}\\ \hline \ea \ \ \ \ \  \stackrel{}{
\longleftrightarrow } \ \ \ \ \ \begin{array}{|c|} \hline {\rm
periodic}\\ V^{(PTHS)}(x)
\\
{\rm on \ an \ arch\ }x(t),\\
{\rm in\ }{\cal J} X \
\cite{hulth}\\
\hline \ea
\\
\\
\ea
 \een
The vertical correspondence originates from the changes of
variables and the horizontal arrows indicate the transition
between the families ${\cal L}$ and ${\cal J}$. One notices the
similarities in the (symmetric or periodic) form of the
functions $V$ as well as in the (straight-line or bent-curve)
shape of their domains $r=r(t) \in l\!\!\!C$ or $x=x(t) \in
l\!\!\!C$.  More details will be provided in our final sect.
\ref{shulth}.

\section{A generalization of the Eckart oscillator \label{sEcka} }

A deeper understanding of one-dimensional systems may be mediated
by an analytic continuation of their real Schr\"{o}dinger equation
 \be
\left [-\,\frac{d^2}{dr^2} + V(r) \right ]
 \, \psi(r) = E  \, \psi(r), \ \ \ \ \ \psi(\pm \infty) = 0.
\label{SE}
  \ee
For one of the simplest particular models $V(x) = \omega\, x^2 +
\lambda\,x^4$ the loss of hermiticity at complex couplings proved
more than compensated by the new insight in its solutions. E.g.,
its spectrum is given by a {\em single} multi-sheeted analytic
function of $\lambda \in l\!\!\! C$ \cite{Wu}. The same idea has
been re-applied to the set of resonances in the cubic well $V(x) =
\omega\, x^2 + \lambda\,x^3$ \cite{Alvarez}. In the cubic case it
was surprising to notice that the spectrum $E_n(\lambda)$ remains
real for the purely imaginary couplings $\lambda = i\,g$. The
rigorous proof of this curious observation dates back to the late
seventies \cite{Caliceti}. It went virtually unnoticed for more
than ten subsequent years.  The phenomenon only re-entered the
physical scene with Zinn-Justin and Bessis who, tentatively,
attributed the strict absence of decay ${\rm Im}\,E_n(i\,g)=0$ to
the mind-boggling real-symmetry-plus-imaginary-antisymmetry of the
cubic force in question \cite{Bessis}.  They also performed a
number of numerical experiments, keeping in mind a paramount
importance of this peculiar symmetry in field theory. There, it
precisely coincides with the fundamental ${\cal PT}$ (i.e.,
parity-plus-time-reversal) invariance. According to Bender et al
\cite{BM} this new type of symmetry might even replace the
traditional requirement of hermiticity in many phenomenological
models.

Within the quantum mechanics itself the parallels between $g\,x^4$
and $i\,g\,x^3$ inspired the numerical and semi-classical study of
the generalized anharmonic forces $V^{(\delta)}(x) = \omega\,x^2+
g\,x^2 (i\,x)^{\delta}$ with a variable real exponent $\delta$
\cite{BB}.  Within the related ${\cal PT}-$symmetric branch of the
``classical" quantum mechanics there appeared new perturbation
series \cite{pert} and quasi-classical approximations \cite{Pham},
a new implementation of supersymmetry \cite{Andrianov} and the new
types of spectra \cite{models}.

Among all the different models with the ${\cal PT}-$symmetrically
broken parity one may distinguish, roughly speaking, between its
``stronger" and ``weaker" violation. The former group is formally
characterized by the globally, asymptotically deformed paths of
integration in eq. (\ref{SE}).  An illustration may be provided by
the elementary ground-state wave function $\psi(x) = \exp \left (
-i\,x^3 + b\,x^2 \right )$ of Bender and Boettcher \cite{BBdva}
which ceases to be integrable on the real line of $x$. The
integrability is only recovered after we bend both the semi-axes
downwards,
 \ben \{x \gg 1 \}
\ \ \longrightarrow \ \ \{ x = \varrho\,e^{-i\,\varphi}\},
 \ \ \ \ \ \ \ \ \ \ \ \ \{x \ll -1
  \} \ \ \longrightarrow \ \ \{ x = - \varrho\,e^{i\,\varphi}\}
  \een
with  $\varrho \gg 1$ and $0<\varphi < \pi/3$. The wave function
obviously corresponds to the quasi-exactly solvable potential
$V(x)=-9x^4-12bix^3+4b^2x^2-6ix$ \cite{BBdva} and mimics the
choice of $\delta=2$ in the family $V^{(\delta)}(x)$. The further
growth of $\delta> 2 $ would make both the asymptotical
$\varphi-$wedges shrink and rotate more and more downwards in the
complex plane.

The second group of the ${\cal PT}$ symmetric examples with a
``weaker" parity breakdown does not leave the real axis of $x$ at
all (i.e., $\varphi \equiv 0$, cf., e.g., \cite{mytri}). This
admits the more natural physical interpretation of the real
physical coordinates. Such a form of the ${\cal P}-$violation has
been also implemented in several numerical and perturbative
models. Their subclass which possesses elementary solutions is
particularly instructive since it incorporates all the so called
shape invariant one-dimensional models of the ordinary quantum
mechanics \cite{shapin}.

In both the groups of examples an overall ${\cal PT}$ symmetry of
the Hamiltonian is, presumably, responsible for its real and
discrete spectrum \cite{BB}. Cannata et al \cite{Cannata} were the
first to notice that one of the various limits $\delta \to \infty$
of the power-law models with $\varphi \to \pi/2-{\cal
O}(1/\delta)$ becomes exactly solvable in terms of Bessel
functions. This re-attracted attention to the related strongly
deformed contours \cite{BBsqw}. More recently, the same merit of
an indirect formal parallel to the Hermitian square well has been
also found for the standard real contour. The related $\delta \to
\infty$ wave functions even degenerated to the Laguerre
polynomials \cite{Morse}.

The latter unexpected emergence of the new exactly solvable model
within the generalized, ${\cal PT}-$symmetric quantum mechanics
encouraged our present study. Indeed, exactly solvable models are
obviously best suited for analyses of methodical questions. In
particular, the class of the ${\cal PT}-$symmetrized shape
invariant oscillators \cite{shapin} does not seem to differ too
much from its Hermitian counterpart. For an explicit analysis of
the details of this correspondence one may simply recall the
numerous explicit formulae available, e.g., in Table 4.1 of the
review \cite{Khare} or in the original factorization
constructions \cite{Infeld} and their Lie-algebraic \cite{Lie},
operator \cite{Dambrowska} or supersymmetric \cite{Levai}
re-interpretations.

Seemingly, one cannot expect any interesting new developments in
the exactly solvable context. Fortunately, in the light of our
recent remark on the spherical harmonic oscillator \cite{PTHO}
non-trivial innovations may be expected in the domain of singular
forces. Indeed, within the ${\cal PT}-$symmetric quantum mechanics
it is possible to {\em avoid} some isolated singularities by a
{\em local} deformation of the integration path. In particular, a
strong repulsion in the origin (so popular in some
phenomenological models \cite{Hall} but fully impenetrable in one
dimension) may be readily controlled by a suitable choice of the
cut.

In the present text we intend to re-attract attention to the
singular forces.  In eq. (\ref{SE}) we shall use the
asymptotically real path of integration which is only {\em
locally} deformed and non-Hermitian. We shall show that this
innovative approach enables us to regularize the one-dimensional
models via their suitable ${\cal PT}$ symmetrization.

\subsection{Eckart model}

Our particular attention will be paid to the exceptional $s-$wave
potential
 \ben
 V^{(Eck)}(x) = \frac{A(A-1)}{\sinh^2 x}-2B\frac{\cosh
x}{\sinh x}
 \een
with the strongly singular core. Usually attributed to Eckart
\cite{Eckart}, this model is solvable on the half-line with $x \in
(0,\infty)$ and, conventionally, $A > 1/2$ and $B>A^2$
\cite{Khare}. Its fixed value of the angular momentum $\ell=0$ is
in effect a non-locality which lowers its practical relevance in
three dimensions. Here, we shall study its ${\cal PT}-$symmetrized
version with the purely imaginary coupling $B = i \beta$. Besides
the obvious relevance of such an exceptional complexified model
with a strong singularity in quantum mechanics, an independent
encouragement of our study is also provided by its obvious
phenomenological and methodical appeal in the context of field
theory, especially in connection with the so called Klauder
phenomenon \cite{Klauder}.

The local deformation of the integration path will enable us to
forget about the strong singularity in the origin. This
deformation will also admit the presence of the so called
irregular components in $\psi(x)\sim x^{1-A}$ near $x=0$. They
would be, of course, unphysical in the usual formalism
\cite{condit}.

\subsection{Solution revisited}

For all these reasons we have to re-analyze the whole
Schr\"{o}dinger equation anew. Our initial choice of the
appropriate variables
 \ben
  \psi(x) = (y-1)^u (y+1)^v \varphi \left (\frac{1-y}{2} \right ),
  \ \ \ \ \ \ \ \ \
y=\frac{\cosh x}{\sinh x}=1-2z
  \een
is still dictated by the arguments of L\'evai \cite{Levai}. Then
we insert $V^{(Eck)}(x) $ in eq. (\ref{SE}) and our change of
variables leads to its new form
\be
z(1-z)\,\varphi''(z) +[c-(a+b+1)z]\,\varphi'(z) -ab\,\varphi(z)=0
 \ee
where
\be
c=1+2u, \ \ \ \ a+b=2u+2v+1, \ \ \ \ ab=(u+v)(u+v+1)+A(1-A)
  \ee
and
\be
4v^2=2B-E, \ \ \ \ \  \ \ \ 4u^2=-2B-E.
  \ee
Our differential equation is of the Gauss hypergeometric type and
its general solution is well known \cite{Ryzhik},
 \be
\varphi(z)= C_1\cdot\ _2F_1(a,b;c;z) + C_2\cdot\ z^{1-c}\
 _2F_1(a+1-c,b+1-c;2-c;z).
 \label{solu}
  \ee
The first thing we notice is that our parameters $a$ and $b$ are
merely functions of the sum $u+v$ and vice versa, $u+v=(a+b-1)/2$.
The immediate insertion then gives the rule $(a-b)^2=(2A-1)^2$ and
we may eliminate
 \be
a = b \pm (2A-1).
 \label{lab}
 \ee
We assume that our solutions obey the standard oscillation
theorems \cite{Hille} and become compatible with the boundary
conditions in eq. (\ref{SE}) at a discrete set of energies, i.e.,
if and only if the infinite series $_2F_1$ terminate. Due to the
complete $a \leftrightarrow b$ symmetry, we only have to
distinguish between the two possible choices of $C_2=0$  and
$C_1=0$.

In the former case with the convenient $b = -N$ (= non-positive
integer) the resulting numbers $a+b$ and $u+v$ prove both real.
Using the definition of $B$ the difference $u-v=-i\beta/(u+v)$
comes out purely imaginary. The related terminating wave function
series (\ref{solu}), i.e.,
 \be
\psi(x) =
 \left ( \frac{1}{\sinh x} \right )^{u+v} \ e^{(v-u)x}
  \cdot\,
\varphi[z(x)]
 \label{soluc1}
  \ee
is asymptotically normalizable if and only if $u+v>0$. This
condition fixes the sign in eq. (\ref{lab}) and gives the explicit
values of all the necessary parameters,
 \be
 a=2A-N-1, \ \ \ \ u+v=A-N-1, \ \ \ \ u-v=-i\,\frac{\beta}{A-N-1}.
 \label{proto}
 \ee
For all the non-negative integers $N \leq N_{max}< A-1$ the
spectrum of energies is obtained in the following closed form,
 \be
 E = -\frac{1}{2}\,\left (u^2+v^2\right ) =
 -\left ( A-N-1 \right )^2 + \frac{\beta^2}{(A-N-1)^2},
 \ \ \ \ \ \ N = 0, 1, \ldots, N_{max}.
 \ee
The normalizable wave functions become proportional to Jacobi
polynomials,
 \be
\varphi[z(x)] = const.\cdot P^{(u/2,v/2)}_N(\coth x).
\label{polynom}
 \ee
Before we start a more thorough discussion of this result we have
shortly to return to the second option with $C_1=0$ in eq.
(\ref{solu}). Curiously enough, this does not bring us anything
new. Although the second Gauss series terminates at the different
$b=c-1 -N$, the factor $z^{1-c}$ changes the asymptotics and one
only reproduces the former solution. All the differences prove
purely formal. In the language of our formulae one just replaces
$u$ by $-u$ in (and only in) {\em both} equations (\ref{soluc1})
and (\ref{proto}). No change occurs in the polynomial
(\ref{polynom}).

\subsection{Spectrum}

The new spectrum of energies seems
phenomenologically appealing. The separate $N-$th energy remains
negative if and only if the imaginary coupling stays sufficiently
weak, $\beta^2 < (A-N-1)^4$. Vice versa, the highest energies may
become positive, with $E=E(N_{max})$ growing extremely quickly
whenever the value of the coupling $A$ approaches its integer
lower estimate $1+N_{max}$ from above. In this way, even a weak
${\cal PT}$ symmetric force $V^{(Eck)}(x)$ is able to produce a
high-lying normalizable excitation. This feature does not seem
connected to the presence of the singularity as it closely
parallels the similar phenomenon observed for the ${\cal PT}$
symmetric Rosen-Morse oscillator which remains regular in the
origin \cite{shapin}. Also, in a way resembling harmonic
oscillators the distance of levels in our model is safely bounded
from below. Abbreviating $D=A-N-1=A_{effective}>0$ its easy
estimate
 \ben
 E_N-E_{N-1}=(2D+1) \left (1 + \frac{\beta^2}{D^2(D+1)^2} \right )
 > 1 \een
(useful, say, in perturbative considerations) may readily be
improved to $ E_N-E_{N-1}>\beta^2/D^2$ at small $D\ll 1$, to $
E_N-E_{N-1}>2D$ at large $D\gg 1$ and, in general, to an algebraic
precise estimate obtainable, say, via MAPLE \cite{Maple}.

Let us emphasize in the conclusion that the formulae we obtained
are completely different from the usual Hermitian $s-$wave results
as derived, say, by L\'evai \cite{Levai}. He has to start from the
regularity in the origin which implies an opposite sign in eq.
(\ref{lab}). This must end up with the constraint $B> 0$.
Moreover, the size of $B$ would limit the number of bound states.
In the present ${\cal PT}$ symmetric setting, a few paradoxes
emerge in this comparison. Some of them may be directly related to
the repulsive real core in our $V^{(Eck)}(x)$ with imaginary $B$.
Thus, one may notice that the {\em increase} of the real repulsion
{\em lowers} the $N-$th energy. In connection with that, the
number of levels {\em grows} with the increase of coupling $A$. In
effect, the new bound-state levels emerge as decreasing from the
positive infinity (!). At the same time, the presence of the
imaginary $B = i\beta$ shifts the whole spectrum upwards precisely
in the manner known from the non-singular models.

\section{A generalization of the P\"{o}schl-Teller potential
 \label{ssPTP} }

\noindent Among all the exactly solvable models in quantum
mechanics the one-di\-men\-si\-o\-nal Schr\"{o}dinger equation
(\ref{SE})
with one of the most elementary bell-shaped
potentials $V^{(bs)}(r) = G/{\cosh^2 r}$ is particularly useful.
Its applications range from the analyses of stability and
quantization of solitons \cite{Bullough} to phenomenological
studies in atomic and molecular physics \cite{physaa}, chemistry
\cite{physbb}, biophysics \cite{physcc} and astrophysics
\cite{physdd}.  Its appeal involves the solvability by different
methods \cite{Levai} as well as a remarkable role in the
scattering \cite{Newton}. Its bound-state wave functions
represented by Jacobi polynomials offer one of the most elementary
illustrations of properties of the so called shape invariant
systems \cite{Khare}. The force $V^{(bs)}(r)$ is encountered in
the so called ${\cal PT}$ symmetric quantum mechanics
\cite{Bender} where it appears as a Hermitian super-partner of a
complex ``scarf" model \cite{Bagchi}.

Curiously enough, it is not too difficult to extend the exact
solvability of the potential $V^{(bs)}(r) $ to all its ``spiked"
(often called P\"{o}schl-Teller \cite{Poeschl}) shape invariant
generalizations
 \be
V^{(PT)}(r)
= -\frac{A(A+1)}{\cosh^2 r} +\frac{B(B-1)}{\sinh^2 r}.
\label{sPTP}
  \ee
In a way resembling the preceding section,
our new one-dimensional Schr\"{o}dinger eq.
(\ref{SE}) is also too singular at $B(B-1) \neq 0$. The
 force $V^{(PT)}(r) $ must be confined to the semi-axis,
  $ r \in (0, \infty)$. This makes the ``improved"
P\"{o}schl-Teller model (\ref{sPTP}) much less useful in practice
since its higher partial waves are not solvable.

\subsection{Regularization}

We may repeat that the impossibility
of using eq. (\ref{sPTP}) in three dimensions (or on the whole
axis in one dimension at least) is felt unfortunate because the
singular potentials themselves are frequently needed in methodical
considerations \cite{Klauder} and in perturbation theory
\cite{Harrell}. They are encountered in phenomenological models
\cite{Sotona} and in explicit computations \cite{Hall} but not too
many of them are solvable \cite{Mathieu}.

We feel mainly inspired by the pioneering letter \cite{BB} where Bender
and Boettcher modified the harmonic oscillator $V^{(HO)}(r) = r^2$
by a
complex downward shift of its axis of coordinates,
 \be
r = x-i\varepsilon, \ \ \ \ \ \  \ x \in (-\infty,\infty).
\label{shov}  \ee The ${\cal PT}$ symmetry of their model
$V^{(BB)}(x)=V^{(HO)}(x-ic) = x^2 - 2icx - c^2$ means its
invariance with respect to the simultaneous reflection $x \to -x$
and complex conjugation $i \to -i$.
Various other complex interactions have been subsequently
 tested and studied
within this framework.

Our study \cite{PTHO} of the
three-dimensional ${\cal PT}$ symmetric harmonic oscillator
offers
the details of our present
 key idea. The shift
(\ref{shov}) has been employed there as a source of a {\em
regularization} of the strongly singular centrifugal term. As long
as $1/(x- i\varepsilon)^2 = (x+i\varepsilon)^2 /
(x^2+\varepsilon^2)^2 $ at any $\varepsilon \neq 0$, this term
remains nicely bounded in a way which is uniform with respect to
$x$. Without any difficulties one may work with $V^{(RHO)}(x) =
r^2(x) +\ell(\ell+ 1)/ r^2(x)$ on the whole real line of $x$. In
what follows the same idea will be applied to the regularized
P\"{o}schl-Teller-like potential
 \ben V^{(RPT)}(x) = V^{(PT)}(x -
i\varepsilon), \ \ \ \ \ \ \ 0 < \varepsilon <\pi/2.
  \een
This potential is a simple function of the L\'{e}vai's
\cite{Levai} variable $g(r)=\cosh 2r$. As long as $g(x -
i\,\varepsilon) = \cosh 2x\,\cos 2\varepsilon - i\,\sinh 2x\,\sin
2\varepsilon$, the new force is ${\cal PT}$ symmetric on the real
line of $x \in (-\infty, \infty)$,
  \ben V^{(RPT)}(-x)= [V^{(RPT)}(x)]^*.
  \een
Due to the estimates $|\sinh^2(x-i\varepsilon)|^2 = \sinh^2 x
\cos^2 \varepsilon +\cosh^2 x \sin^2 \varepsilon = \sinh^2 x
+\sin^2 \varepsilon$ and $|\cosh^2(x-i\varepsilon)|^2 = \sinh^2 x
+\cos^2 \varepsilon$ the regularity of $V^{(RPT)}(x)$ is
guaranteed for all its parameters $\varepsilon \in (0, \pi/2)$.

\subsection{Solutions}

In a way paralleling the three-dimensional oscillator the mere
analytic continuation of the $s-$wave bound states does not give
the complete solution.  One must return to the original
differential equation (\ref{SE}). There we may conveniently fix $A
+1/2=\alpha>0$ and $B-1/2=  \beta>0$ and write
 \be
\left (-\,\frac{d^2}{dx^2} +
\frac{\beta^2-1/4}{\sinh^2 r(x)}
 -\frac{\alpha^2-1/4}{\cosh^2 r(x)}  \right )
 \, \psi(x) = E  \, \psi(x), \ \ \ \ \ \ r(x)= x-i\varepsilon.
\label{SEb}
 \ee
This is the Gauss differential equation
 \be
z(1+z)\,\varphi''(z) +[c+(a+b+1)z]\,\varphi'(z)
+ab\,\varphi(z)=0
\label{gauss}
 \ee
in the new variables
 \ben
\psi(x) = z^\mu(1+z)^\nu\varphi(z),
\ \ \ \ \ \ \ \ z = \sinh^2r(x)
 \een
using the suitable re-parameterizations
 \ben
\alpha^2=(2\nu-1/2)^2,
\ \ \ \ \ \ \ \ \
 \beta^2=(2\mu-1/2)^2,
\ \ \ \ \ \ \ \ \
 \een
 \ben
2\mu+1/2=c, \ \ \ \ \
2\mu+2\nu=a+b, \ \ \ \ \
E= -(a-b)^2.
 \een
In the new notation we have the wave functions
 \be
\psi(x) = \sinh^{\tau \beta+1/2}[r(x)]
 \cosh^{\sigma\alpha+1/2}[r(x)]\,\varphi[z(x)]
\label{formula}
  \ee
with the sign ambiguities $\tau = \pm 1$ and $\sigma=\pm 1$ in
$2\mu=\tau \beta+1/2$ and $2\nu=\sigma\alpha+1/2$. This formula
contains the general solution of hypergeometric eq. (\ref{gauss}),
 \be
\varphi(z) = C_1\ _2F_1(a,b;c;-z) + C_2z^{1-c}\
_2F_1(a+1-c,b+1-c;2-c;-z). \label{gensol}
  \ee
The solution should obey the complex version of the
Sturm-Liouville oscillation theorem \cite{Hille}.  In the case of
the discrete spectra this means that we have to demand the
termination of our infinite hypergeometric series. This suppresses
an asymptotic growth of $\psi(x)$ at $x\to\pm\infty$.

In a deeper analysis let us first put $C_2=0$. We may satisfy the
termination condition by the non-positive integer choice of
$b=-N$.  This implies that $a=N+1 +\sigma\alpha +\tau  \beta$ is
real and that our wave function may be made asymptotically
(exponentially) vanishing under certain conditions. Inspection of
the formula (\ref{formula}) recovers that the boundary condition
$\psi(\pm \infty) = 0$ will be satisfied if and only if
  \ben 1 \leq 2N+1\leq 2N_{max}+1<-\sigma
\alpha - \tau  \beta.
 \een The closed Jacobi polynomial
representation of the wave functions follows easily,
  \ben
\varphi[z(x)] =C_1\ \frac{N!\Gamma(1+\tau \beta)}{\Gamma(N+1+\tau
\beta)} \ P_N^{(\tau \beta,\sigma\alpha)}[\cosh 2r(x)].
  \een
The final insertions of parameters define the spectrum of
energies,
 \be
E=-( 2N+1+\sigma \alpha + \tau  \beta)^2 < 0. \label{energy}
 \ee
Now we have to return to eq.  (\ref{gensol}) once more. A careful
analysis of the other possibility $C_1=0$ does not recover
anything new.  The same solution is obtained, with $\tau$ replaced
by $-\tau$.  We may keep $C_2=0$ and mark the two independent
solutions by the sign $\tau$. Once we define the maximal integers
$N_{max}^{(\sigma,\tau)}$ which are compatible with the inequality
 \be
2N_{max}^{(\sigma,\tau)}+1< -\sigma\alpha-\tau \beta \label{maxes}
 \ee
we get the constraint $N \leq N_{max}^{(\sigma,\tau)}$. The set of
our main quantum numbers is finite.

\subsection{Paradoxes}

Let us now compare our final result (\ref{energy}) with the known
$\varepsilon=0$ formulae for $s$ waves \cite{Levai}. An additional
physical boundary condition must be imposed in the latter singular
limit \cite{conditab}.  This condition fixes the unique pair
$\sigma = -1$ and $\tau = +1$. Thus, the set of the $s-$wave
energy levels $E_N$ is not empty if and only if $\alpha- \beta>
1$. In contrast, all our $\varepsilon > 0$ potentials acquire a
uniform bound $|V^{(RPT}(x)| < const < \infty$. Due to their
regularity, no additional constraint is needed. Our new spectrum
$E^{(\sigma,\tau)}_N$ becomes richer. For the sufficiently strong
couplings it proves composed of the three separate parts,
 \ben
E^{(-,-)}_N< 0, \ \ \ \ \ 0 \leq N \leq N_{max}^{(-,-)}, \ \ \ \ \
\ \alpha+ \beta > 1,
 \een
 \be
E^{(-,+)}_N<0, \ \ \ \ \ \ 0 \leq  N \leq N_{max}^{(-,+)}, \ \ \ \
\ \ \alpha> \beta + 1,
 \ee
 \ben
E^{(+,-)}_N<0, \ \ \ \ \ \ 0 \leq  N \leq N_{max}^{(+,-)},
 \ \ \ \ \ \
\beta > \alpha+1.
 \een
The former one is non-empty at $ A + B > 1$ (with our above
separate conventions $A > -1/2$ and $B > 1/2$). Concerning the
latter two alternative sets, they may exist either at  $A> B$ or
at $B > A+2$, respectively. We may summarize that in a parallel to
the ${\cal PT}$ symmetrized harmonic oscillator of ref.
\cite{PTHO} we have the $N_{max}^{(-,+)}+1$ quasi-odd or
``perturbed", analytically continued $s-$wave states (with a nodal
zero near the origin) complemented by certain additional
solutions.

In the first failure of a complete analogy the number
$N^{(-,-)}_{max}+1$ of our quasi-even states proves systematically
higher than $N^{(-,+)}_{max}+1$, especially at the larger
``repulsion" $ \beta \gg 1$. This is a certain paradox,
strengthened by the existence of another quasi-odd family which
behaves very non-perturbatively.  Its members (with the ground
state $\psi_0^{(+,-)}(x) =\cosh^{A+1} [r(x)]\sinh^{1-B} [r(x)]$
etc) do not seem to have any $s-$wave analogue. They are formed at
the prevalent repulsion $B>A+2$ which is even more
counter-intuitive. The exact solvability of our example enables us
to understand this apparent paradox clearly. In a way
characteristic for many ${\cal PT}$ symmetric systems some of the
states are bound by an antisymmetric imaginary well. The whole
history of the ${\cal PT}$ symmetric models starts from the purely
imaginary cubic force \cite{Bessis} after all. A successful
description of its perturbative forms $V(x) = \omega x^2+i\lambda
\,x^3$ is not so enigmatic \cite{Caliceti}, especially due to its
analogies with the real and symmetric $V(x) = \omega x^2+
\lambda\,x^4$ \cite{Alvarez}. The similar mechanism creates the
states with $(\sigma,\tau)=(+,-)$ in the present example. A
significant novelty of our new model $V^{(RPT)}(x)$ lies in the
dominance of its imaginary component {\em at the short distances},
$x \approx 0$. Indeed, we may expand our force to the first order
in the small $\varepsilon>0$. This gives the approximation
 \be
\frac{1}{\sinh^2(x-i\varepsilon)}=
\frac{\sinh^2(x+i\varepsilon)}{(\sinh^2x + \sin^2\varepsilon)^2}
=\frac{1}{\sinh^2x}+2i\varepsilon \frac{\cosh x}{\sinh^3 x}
+{\cal O}(\varepsilon^2).
 \label{sini}
 \ee
We see immediately the clear prevalence of the imaginary part at
the short distances, especially at all the negligible $A = {\cal
O}(\varepsilon^2)$.

An alternative approach to the above paradox may be mediated by a
sudden transition from the domain of a small $\varepsilon \approx
0$ to the opposite extreme with $\varepsilon \approx \pi/2$. This
is a shift which changes $\cosh x$ into $\sinh x$ and vice versa.
It intertwines the role of $\alpha$ and $ \beta$ as a strength of
the smooth attraction and of the singular repulsion, respectively.
The perturbative/non-perturbative interpretation of both our
quasi-odd subsets of states becomes mutually interchanged near
both the extremes of $\varepsilon$.

The dominant part (\ref{sini}) of our present model leaves its
asymptotics comparatively irrelevant. In contrast to many other
${\cal PT}$ symmetric models as available in the current
literature our potential vanishes asymptotically,
 \ben V^{(RPT)}(x) \to 0, \ \ \ \ \ \ \ \ x \to \pm \infty.
  \een
An introduction and analysis of continuous spectra in the ${\cal
PT}$ symmetric quantum mechanics seems rendered possible at
positive energies. This question will be left open here. In the
same spirit of a concluding remark we may also touch the problem
of the possible breakdown of the ${\cal PT}$ symmetry. This has
recently been studied on the background of the supersymmetric
quantum mechanics \cite{Andrianov}. In our present solvable example
the violation of the ${\cal PT}$ symmetry is easily mimicked by
the complex choice of the couplings $\alpha$ and $ \beta$. Due to
our closed formulae the energies will still stay real, provided
only that ${\rm Im}\ (\sigma\alpha+ \tau \beta)=0$.

\section{  ${\cal PT}$ symmetric Hulth\'en potential
\label{shulth}
}

Quantum mechanics is often forced to formulate its predictions
numerically. Its exactly solvable models are scarce. Only their
subclass in one dimension is broader and, in this sense,
privileged and exceptional. It involves the harmonic oscillator
and Morse potentials (with bound states expressible in terms of
the Laguerre polynomials) as well as several other, less popular
models solvable in terms of the polynomials of Jacobi etc (cf.,
e.g., review  \cite{Khare} for more details).

One encounters a high degree of analogy between the
real and complex forces for the exactly solvable models based on
the use of Jacobi polynomials. An appropriate exactly solvable
${\cal PT}$ symmetrization of both the one-dimensional
shape-invariant models with the ``canonical" Rosen-Morse and
scarf-hyperbolic shapes has been recently proposed and analyzed
in refs. \cite{Bagchi}.

As we already noticed, a word of warning comes from the
phenomenologically most appealing Coulomb interaction. Up to
now, its only available ${\cal PT}$ symmetric simulation proves
merely partially solvable \cite{coho}.  In this section
we intend to offer a partial remedy.  We shall
derive and describe a complete and exact solution for an
appropriate ${\cal PT}$ symmetric complexification of the
singular phenomenological Hulth\'{e}n potential \cite{Newton}
which is known to mimic very well the shape of the Coulombic
force in the vicinity of its singularity.

\subsection{Method: The Liouvillean change of variables}

In the first step let us recollect that in the spirit of the old
Liouville's paper \cite{Liouville} the change of the (real)
coordinates (say, $r \leftrightarrow \xi$) in Schr\"{o}dinger
equation
 \be
\left[-\,\frac{d^2}{dr^2} + W(r)\right]\, \chi(r) = -\kappa^2
\,\chi(r)
 \label{SEor}
 \ee
may sometimes mediate a transition between two different
potentials.  It is easy to show \cite{Olver} that once we forget
about boundary conditions one simply has to demand the existence
of the invertible function $r=r(\xi)$ and its few derivatives
$r'(\xi), \, r''(\xi), \ldots$ in order to get the explicit
correspondence between the two bound state problems, viz.,
original (\ref{SEor}) and the new Schr\"{o}dinger equation
with the known wave functions
 \be
\Psi(\xi)={ \chi[r(\xi)] \over \sqrt{r'(\xi)}} \label{trenky}
 \ee
generated by the new interaction
\be
 V(\xi)-E=\left [
 r'(\xi)
 \right ]^2
 \left \{
 W[r(\xi)]+\kappa^2
 \right \}
+
 \frac{3}{4}
 \left [
 {r''(\xi) \over r'(\xi)}
 \right ]^2
 -
 \frac{1}{2}
 \left [
 {r'''(\xi) \over r'(\xi)}
 \right ].
 \label{newpot}
 \ee
One might recall the well known mapping between the Morse and
harmonic Laguerre-related oscillators as one of the best known
explicit illustrations. For it, the necessary preservation of
the correct physical boundary conditions is very straightforward
to check \cite{Newton}.

In the Jacobi-polynomial context the Liouvillean changes of
variables have been applied systematically to all the Hermitian
models (cf. Figure 5.1 in the review \cite{Khare} or ref.
\cite{Varshni} for a more detailed illustration).  A similar
thorough study is still missing for the ${\cal PT}$ symmetric
models within the same class.

In the present letter we shall try to fill the gap.  For the
sake of brevity we shall only restrict our attention to the
${\cal PT}$ symmetric initial eq. (\ref{SEor}) with the
P\"{o}schl-Teller potential studied and solved exactly in our
recent preprint \cite{PTP},
 \be
 W(r)=\frac{\beta^2-1/4}{\sinh^2 r}
 -\frac{\alpha^2-1/4}{\cosh^2 r}, \ \ \ \ \ \ r = x -
 i\varepsilon,
 \ \ \ \ \ \ \ \ \ x \in (-\infty, \infty)
\label{PTex}
  \ee
The normalizable solutions are proportional to the Jacobi
polynomials,
 \ben
\chi(r) = \sinh^{\tau \beta+1/2}r
 \cosh^{\sigma\alpha+1/2}r\,\
  P_n^{(\tau \beta,\sigma\alpha)}(\cosh 2r)
  \een
at all the negative energies $-\kappa^2 < 0$ such that
 \ben
 \kappa=
 \kappa^{(\sigma,\tau)}_n=-\sigma\alpha-\tau\beta -2n-1>0.
 \een
These bound states are numbered by $n = 0,1,\ldots, n_{max}^{
(\sigma,\tau)}$ and by the generalized parities $\sigma=\pm 1$
and $ \tau =\pm 1$.

We may note that our initial ${\cal PT}$ symmetric model
(\ref{SEor}) remains manifestly regular provided only that its
constant downward shift of the coordinates $r = r_{(x)} = x -
i\,\varepsilon$ remains constrained to a finite interval,
$\varepsilon \in (0,\pi/2)$.  In this sense our initial model
(\ref{PTex}) is closely similar to the shifted harmonic
oscillator.  At the same time, one still misses an analogue of a
transition to its Morse-like final partner $V(\xi)$. In a key step
of its present construction let us first pick up the following
specific change of the axis of coordinates,
 \be
 \sinh r_{(x)} (\xi)= - i e^{i\xi}, \ \ \ \ \ \ \ \ \xi = v - iu.
\label{tren}
 \ee
The main motivation of such a tentative assignment lies in the
related shift and removal of the singularity (sitting at $r=0$)
to infinity ($u \to +\infty$). In fact, one cannot proceed
sufficiently easily in an opposite direction, i.e., from a
choice of a realistic $V(\xi)$ to a re-constructed $r(\xi)$.
This is due to the definition (\ref{newpot}) containing the
third derivatives and, hence, too complicated to solve.

We shall see below that we are quite lucky with our purely trial
and error choice of eq. (\ref{tren}). Firstly, we already clearly
see that the real line of $x$ becomes mapped upon a manifestly
${\cal PT}$ symmetric curve $\xi = v - iu$ in accordance with the
compact and invertible trigonometric rules
 \ben
 \ba
 \sinh x \cos \varepsilon = e^u\,\sin v,\\
 \cosh x \sin \varepsilon = e^u\,\cos v,
 \ea
 \een
i.e., in such a way that
 \ben
 \ba
 v =\arctan \left (
\frac{\tanh x}{\tan \varepsilon} \right )= v_{(x)} \in \left
(v_{(-\infty)}, v_{(\infty)}\right )\equiv \left
(-\frac{\pi}{2}+\varepsilon, \frac{\pi}{2}-\varepsilon \right ),\\
u = u_{(x)} = \frac{1}{2} \ln \left ( \sinh^2x+\sin^2\varepsilon
 \right ) .
 \ea
 \een
This relationship is not too different from the Morse-harmonic
equivalence studied in ref. \cite{Morse}. Our present path of
$\xi $ is a very similar down-bent arch which starts in its left
imaginary minus infinity, ends in its right imaginary minus
infinity while its top lies at $x=v=0$ and $-u=-u_{(0)}= \ln
1/\sin \varepsilon >0$. The top may move towards the singularity
in a way mimicked by the diminishing shift $\varepsilon \to 0$.
Indeed, although the singularity originally occurred at the
finite value $r\to 0$, it has now been removed upwards, i.e., in
the direction of $-u \to +\infty$.

\subsection{Consequences}

The first consequence of our particular change of variables
(\ref{tren}) is that it does not change the asymptotics of the
wave functions. As long as $r'(\xi)= i\tanh r(\xi)$ the
transition from eq. (\ref{SEor}) to (\ref{SE}) introduces just
an inessential phase factor in $\Psi(\xi)$. This implies that
the normalizability (at a physical energy) as well as its
violations (off the discrete spectrum) are both in a one-to-one
correspondence.

The explicit relation between the old and new energies and
couplings is not too complicated. Patient computations reveal
its closed form. With a bit of luck, the solution proves
non-numerical.  The new form of the potential and of its binding
energies is derived by the mere insertion in eq. (\ref{newpot}),
 \be
 V(\xi)=
 \frac{A}{(1-e^{2i\xi})^2}+\frac{B}{1-e^{2i\xi}},
\ \ \ \ \ \  E=\kappa^2.
 \label{hulth}
 \ee
At the imaginary $\xi$ and vanishing $A=0$ this interaction
coincides with the Hulth\'en potential.

In the new formula for the energies one has to notice their
positivity. This is extremely interesting since the potential
itself is asymptotically vanishing at both ends of its
integration path.  One may immediately recollect that a similar
paradox has already been observed in a few other ${\cal PT}$
symmetric models with an asymptotic decrease of the potential to
minus infinity \cite{BBdva,Sesma}.

The exact solvability of our modified Hulth\'en potential is not
yet guaranteed at all. A critical point is that the new couplings
depend on the old energies and, hence, on the discrete quantum
numbers $n$, $\sigma$ and $\tau$ in principle.  This could
induce an undesirable state-dependence into our new potential.
Vice versa, the closed solvability of the constraint which
forbids this state-dependence will be equivalent to the
solvability at last.

A removal of the obstacle means in effect a transfer of the
state-dependence (i.e., of the $n-$, $\sigma-$ and
$\tau-$dependence) in
 \ben A=A(\alpha) = 1 - \alpha^2, \ \ \ \ \ \
\ \ C \ (= A +B) =\kappa^2-\beta^2 \een from $C$ to $\beta$.  To
this end, employing the known explicit form of $\kappa$ we may
re-write
\be
C=C(\sigma,\tau,n)= (\sigma\alpha+2n+1) (\sigma\alpha+2n+1
+2\tau \beta).
\ee
This formula is linear in $\tau\beta$ and, hence, its inversion
is easy and defines the desirable state-dependent quantity
$\beta=\beta(\sigma,\tau,n)$ as an elementary function of the
constant $C$.  The new energy spectrum acquires the closed form
\be
E=E(\sigma,\tau,n)=A+B+\frac{1}{4}\,
\left [
\sigma\alpha+2n+1-\frac{A+B}{\sigma\alpha+2n+1}
\right ]^2.
\ee
Our construction is complete.  The range of the quantum numbers
$n, \ \sigma$ and $\tau$ remain the same as above.

In the light of our new result we may now split the whole family
of the exactly solvable ${\cal PT}$ symmetric models in the two
distinct categories.  The first one ``lives" on the real line
and may be represented or illustrated not only by the popular
Laguerre-solvable harmonic oscillator  \cite{PTHO} but also by
our initial P\"{o}schl-Teller Jacobi-solvable force
(\ref{PTex}).

The second category requires a narrow arch-shaped path of
integration which all lies confined within a vertical strip.  It
contains again both the Laguerre and Jacobi solutions.  The
former ones may be represented by the complex Morse model of
ref. \cite{Morse}.  Our present new Hulth\'en example offers its
Jacobi solvable counterpart.  The scheme becomes, in a way,
complete.

The less formal difference between the two categories may be
also sought in their immediate physical relevance.  Applications
of the former class may be facilitated by a limiting transition
which is able to return them back on the usual real line.  In
contrast, the second category may rather find its most useful
place in the methodical considerations concerning, e.g., field
theories and parity breaking \cite{Bentri}.  Within the quantum
mechanics itself an alternative approach to the second category
might also parallel studies \cite{Cannata} of the ``smoothed"
square wells in non-Hermitian setting.

In the conclusion let us recollect that the ${\cal PT}$ symmetry
of a Hamiltonian replaces and, in a way, {generalizes} its usual
hermiticity. This is the main reason why there existed a space
for a new solvable model among the singular interactions.  An
exactly solvable example with an ``intermediate" (i.e.,
hyperbola-shaped) arc of coordinates remains still to be
discovered.  Indeed, this type of a deformed contour has only
been encountered in the ``quasi-solvable" (i.e., partially
numerical) model of ref. \cite{BBdva}) and in the general
unsolvable forces studied by several authors by means of the
perturbative \cite{Caliceti}, numerical \cite{Guardiola} and WKB
\cite{WKB} approximative techniques.

\section*{Acknowledgement}

Partially supported by the grant Nr. A 1048004 of the Grant
Agency of the Academy of Sciences of the Czech Republic.


\newpage
\begin{thebibliography}{99}

\bibitem{Khare}
F. Cooper, A. Khare and U. Sukhatme, Phys. Rep. 251 (1995) 267.

\bibitem{Bender}
C. M. Bender, S. Boettcher and P. N. Meisinger,
J. Math. Phys. 40, 2201 (1999).

\bibitem{BG}
Buslaev V and Grechi V 1993 J. Phys. A: Math. Gen. 26 5541

\bibitem{Andrianov}
A. A. Andrianov, M. V. Ioffe, F. Cannata and J. P. Dedonder, Int.
J. Mod. Phys. A 14 (1999) 2675.

\bibitem{Cannata}
F. Cannata, G. Junker and J. Trost, Phys. Lett. { A 246} (1998)
219.

\bibitem{BB}
C. M. Bender and S. Boettcher, Phys. Rev. Lett. { 24} (1998) 5243

\bibitem{Morse}
M. Znojil, Phys. Lett. A. 264 (1999) 108.

\bibitem{PTHO}
M. Znojil, Phys. Lett. A 259 (1999) 220.

\bibitem{coho}
M. Znojil, J. Phys. A: Math. Gen. 32 (1999) 4563.

\bibitem{LevaiZ}
G. Levai and M. Znojil, in preparation.

\bibitem{Bagchi}
B. Bagchi and R. Roychoudhury,
 J. Phys. A: Math. Gen. 33 (2000) L1.

\bibitem{shapin}
M. Znojil, J. Phys. A: Math. Gen. 33 (2000) L61.

\bibitem{Ecka}
M. Znojil, arXiv: quant-ph/9912027, unpublished.

\bibitem{PTP}
M. Znojil, arXiv: quant-ph/9912079, unpublished.

\bibitem{hulth}
M. Znojil, arXiv: math-ph/0002017, unpublished.

\bibitem{Wu}
C. Bender and T. T. Wu, Phys. Rev. Lett. 21 (1968) 406;

B. Simon, Int. J. Quant. Chem. 21 (1982) 3.

\bibitem{Alvarez}
G. Alvarez, J. Phys. A: Math. Gen. 27 (1995) 4589.

\bibitem{Caliceti}
E. Caliceti, S. Graffi and M. Maioli, Commun. Math. Phys. 75
(1980) 51

\bibitem{Bessis}
Daniel Bessis, private communication.

\bibitem{BM}
C. M. Bender and K. A. Milton, Phys. Rev. D 55 (1997) R3255 and D
57 (1998) 3595 and  J. Phys. A: Math. Gen. 32 (1999) L87.

\bibitem{pert}
F. Fern\'andez, R. Guardiola,  J. Ros and M. Znojil,  J. Phys. A:
Math. Gen. 31 (1998) 10105;

C. M. Bender and G. V. Dunne, J. Math. Phys. 40 (1999) 4616.

\bibitem{Pham}
G. Alvarez, J. Phys. A: Math. Gen. 22 (1989) 617;

E. Delabaere and F. Pham, Phys. Lett. A 250 (1998) 25 and 29.

\bibitem{models}
C. M. Bender and A. V. Turbiner, Phys. Lett. { A 173} (1993) 442;

C. M. Bender, G. V. Dunne and P. N. Meisinger, Phys. Lett. A 252
(1999) 272.

\bibitem{BBdva}
C. M. Bender and S. Boettcher, J. Phys.  A: Math. Gen. 31 (1998)
L273.

\bibitem{mytri}
F. Fern\'andez, R. Guardiola,  J. Ros and M. Znojil, J. Phys. {
A}: Math. Gen. 32 (1999) 3105.

\bibitem{BBsqw}
C. M. Bender, S. Boettcher, H.F. Jones and Van M. Savage, J. Phys.
A: Math. Gen. 32 (1999) 6771.

\bibitem{Infeld}
L. Infeld and T. E. Hull, Rev. Mod. Phys. 23 (1951) 21.

\bibitem{Lie}
W. Miller Jr, Lie Theory of Special Functions, AcAdemic, New York,
1968.

\bibitem{Dambrowska}
J. W. Dabrowska, A. Khare and U. Sukhatme, J. Phys. A: Math. Gen.
21 (1988) L195.

\bibitem{Levai}
G. L\'evai, J. Phys. A: Math. Gen. 22 (1989) 689.

\bibitem{Hall}
V. C. Aguilera-Navarro, G. A. Estevez and R. Guardiola, J. Math.
Phys. 31 (1990), 99;

F. M. Fern\'{a}ndez, Phys. Lett. A 160, 511 (1991);

R. Hall and N. Saad, J. Phys. A: Math. Gen. 32, 133 (1999).

\bibitem{Eckart}
C. Eckart, Phys. Rev. 35 (1930) 1303

\bibitem{Klauder}
L. C. Detwiler and J. R. Klauder, Phys. Rev. D 11 (1975) 1436.

\bibitem{condit}
M. Znojil, LANL preprint quant-ph/9811088, to appear in Phys. Rev.
A.

\bibitem{Ryzhik}
M. Abramowitz and I. A. Stegun, Handbook of Mathematical
Functions, NBS, Washington, 1964.

\bibitem{Hille}
E. Hille, Lectures on Ordinary Differential Equations,
Addison-Wesley, Reading, 1969.

\bibitem{Maple}
B. W. Char et al, Maple V, Springer, New York, 1991.


\bibitem{Bullough}
G. B. Whitham, Linear and Nonlinear Waves (John Wiley and Sons,
New York, 1974);

R. Jackiw, Rev. Mod. Phys. 49, 681 (1977).

\bibitem{physaa}
V. I Kukulin, V. M. Krasnopol'sky and J. Hor\'{a}\v{c}ek, Theory
of resonances: Principles and Applications (Kluwer, Dordrecht,
1989);

R. Dutt, A. Gangopadhyaya, C. Rasinarin and U. Sukhatme, Phys.
Rev. A 60, 3482 (1999).

\bibitem{physbb}
 R. P. Bell,
     The Tunnel Effect in Chemistry
      (Chapman and Hall, London, 1980).

\bibitem{physcc}
 D. De Vault, Quantum Mechanical Tunneling in Biological Systems
       (Cambridge University Press, London, 1984).

\bibitem{physdd}
H. R. Beyer, Comm. Math. Phys. 204, 397 (1999).

\bibitem{Newton}
R. G. Newton, Scattering Theory of Waves and Particles
 (Springer Verlag, New York, 1982), p. 438.

\bibitem{Poeschl}
G. P\"{o}schl and E. Teller, Z. Physik 83, 143 (1933);

S. Fl\"{u}gge, Practical Quantum Mechanics I (Springer, Berlin,
1971).

\bibitem{Harrell}
E. M. Harrell, Ann. Phys. (NY) 105, 379 (1977).

\bibitem{Sotona}
M. Sotona and J. \v{Z}ofka, Phys. Rev. C 10, 2646 (1974) and Czech.
J. Phys. B 28, 593 (1978);

R. Dutt and Y. P. Varshni, J. Phys. B: At. Mol. Phys. 20, 2437 (1987);

J. Vacek, K. Konvi\v{c}ka and P. Hobza, Chem. Phys. Lett. 220, 83 (1994).

\bibitem{Mathieu}
A. Kratzer, Z. Physik 3, 289 (1920);

N. A. W. Holzwarth, J. Math. Phys. 14, 191 (1973);

E. Papp, Phys. Lett. A 157, 192 (1991).

\bibitem{conditab}
K. M. Case, Phys. Rev. 80, 797 (1950);

L. D. Landau and E. M. Lifschitz, Quantum Mechanics (Pergamon,
London, 1960), ch. V, par. 35;

W. M. Frank, D. J. Land and R. M. Spector, Rev. Mod. Phys. 43, 36
(1971);

M. Znojil, LANL preprint quant-ph/9811088 and Phys. Rev. A, to
appear.

\bibitem{Liouville}
J. Liouville, J. Math. Pures Appl. 1 (1837) 16;

M. Znojil, J. Phys. A: Math. Gen. 27 (1994) 4945.

\bibitem{Olver}
G. Jaff\'e, Z. Phys. 87 (1933) 535;

F. W. J. Olver, Introduction to Asymptotics and Special Functions
(Academic, New York,  1974), Chap. VI;

G. L\'evai, J. Phys. A: Math. Gen. 22 (1989) 689.

\bibitem{Varshni}
R. Dutt, A. Khare and Y. P. Varshni,
 J. Phys. A: Math. Gen. 28 (1995) L107.

\bibitem{Sesma}
J. Sesma and ...,
preprint of the University of Zaragoza,
 submitted for publication;

G. Alvarez and J. C. ...,
preprint of the Universidad Complutense, Madrid,
 submitted for publication.

\bibitem{Bentri}
C. M. Bender and A. V. Turbiner, Phys. Lett. { A 173} (1993) 442;

C. M. Bender and K. A. Milton, Phys. Rev. D 55 (1997) R3255
 and D
57 (1998) 3595 and  J. Phys. A: Math. Gen. 32 (1999) L87.

C. M. Bender, S. Boettcher, H.F. Jones and Van M. Savage, J. Phys.
A: Math. Gen. 32 (1999) 6771.

\bibitem{Guardiola}
M. Znojil, J. Phys.  A: Math. Gen. 32 (1999) 7419.

\bibitem{WKB}
A. Voros, J. Phys.  A: Math. Gen. 27 (1994) 4653

\end{thebibliography}
\end{document}